\documentclass[aps,pre,groupedaddress,twocolumn,10pt, reprint,superscriptaddress, longbibliography]{revtex4-2}

\usepackage{amsmath}
\usepackage{amssymb}
\usepackage{amsthm}
\usepackage{bbm}
\usepackage{graphicx}
\usepackage{amsmath}
\usepackage{graphicx}
\usepackage{color}
\usepackage{upgreek}
\usepackage[linkcolor = blue, citecolor = blue, urlcolor = blue, colorlinks = true]{hyperref}
\usepackage{amssymb}
\usepackage{bm}
\usepackage[capitalise]{cleveref}
\usepackage{textcomp}
\usepackage{hyperref}
\usepackage[usenames,dvipsnames]{xcolor}
\usepackage[normalem]{ulem}
\usepackage{wasysym}

\usepackage{siunitx}
\graphicspath{{./fig_final/}}

\newcommand{\Ecoli}{{\it E.~coli}}

\newcommand\diff{\mathrm{d}}
\renewcommand{\vec}[1]{\mathbf{#1}}
\renewcommand{\imath}[0]{\mathsf{i}}

\hypersetup{colorlinks=true, linkcolor=blue!50!black, urlcolor=blue!50!black, citecolor=blue!50!black}

\definecolor{ABpurple}{RGB}{128, 0, 128}
\definecolor{ABred}{RGB}{255, 0, 0}
\definecolor{ABgreen}{RGB}{0, 255, 0}
\definecolor{ABbrown}{RGB}{128, 64, 0}
\definecolor{ABblue}{RGB}{0, 0, 255}

\begin{document}
\title{Quantitative characterization of run-and-tumble statistics in bulk bacterial suspensions}
\author{Yongfeng Zhao}
\thanks{Y.Z. and C.K. contributed equally.}
\email{yfzhao2021@suda.edu.cn}
\affiliation{Center for Soft Condensed Matter Physics and Interdisciplinary Research \& School of Physical Science and Technology, Soochow University, Suzhou 215006, China}
\affiliation{School of Physics and Astronomy and Institute of Natural Sciences, Shanghai Jiao Tong University, Shanghai 200240, China}
\affiliation{School of Biomedical Sciences, Li Ka Shing Faculty of Medicine, University of Hong Kong, Pok
Fu Lam, Hong Kong, PR China}
\affiliation{Universit\'e de Paris, MSC, UMR 7057 CNRS, 75205 Paris, France}
\author{Christina Kurzthaler}
\thanks{Y.Z. and C.K. contributed equally.}
\email{ckurzthaler@pks.mpg.de}
\affiliation{Max Planck Institute for the Physics of Complex Systems, 01187 Dresden, Germany}
\affiliation{Center for Systems Biology Dresden, 01307 Dresden, Germany}
\affiliation{Department of Mechanical and Aerospace Engineering, Princeton University, Princeton, New Jersey 08544, USA}
\affiliation{Institut f\"ur Theoretische Physik, Universit\"at Innsbruck, Technikerstra{\ss}e 21A, A-6020 Innsbruck, Austria}
\author{Nan Zhou}
\affiliation{ZJU-Hangzhou Global Scientific and Technological Innovation Center, Zhejiang University, Hangzhou  311200, China}
\author{Jana Schwarz-Linek}
\affiliation{School of Physics and Astronomy, University of Edinburgh, James Clerk Maxwell Building, Peter Guthrie Tait Road, Edinburgh EH9 3FD, United Kingdom}
\author{Clemence Devailly}
\affiliation{School of Physics and Astronomy, University of Edinburgh, James Clerk Maxwell Building, Peter Guthrie Tait Road, Edinburgh EH9 3FD, United Kingdom}
\author{Jochen Arlt}
\affiliation{School of Physics and Astronomy, University of Edinburgh, James Clerk Maxwell Building, Peter Guthrie Tait Road, Edinburgh EH9 3FD, United Kingdom}
\author{Jian-Dong Huang}
\affiliation{School of Biomedical Sciences, Li Ka Shing Faculty of Medicine, University of Hong Kong, Pok Fu Lam, Hong Kong, PR China}
\affiliation{CAS Key Laboratory of Quantitative Engineering Biology, Shenzhen Institute of Synthetic Biology, Shenzhen Institutes of Advanced Technology, Chinese Academy of Sciences, Shenzhen 518055, China}
\author{Wilson C. K. Poon}
\affiliation{School of Physics and Astronomy, University of Edinburgh, James Clerk Maxwell Building, Peter Guthrie Tait Road, Edinburgh EH9 3FD, United Kingdom}
\author{Thomas Franosch}
\affiliation{Institut f\"ur Theoretische Physik, Universit\"at Innsbruck, Technikerstra{\ss}e 21A, A-6020 Innsbruck, Austria}
\author{Vincent A. Martinez}
\affiliation{School of Physics and Astronomy, University of Edinburgh, James Clerk Maxwell Building, Peter Guthrie Tait Road, Edinburgh EH9 3FD, United Kingdom}
\author{Julien Tailleur}
\email{julien.tailleur@univ-paris-diderot.fr}
\affiliation{Universit\'e de Paris, MSC, UMR 7057 CNRS, 75205 Paris, France}
\date{\today}

\begin{abstract}
We introduce a numerical method to extract the parameters of run-and-tumble dynamics from  experimental measurements of the intermediate scattering function. We show that proceeding in Laplace space is unpractical and employ instead renewal processes to work directly in real time. \if{We introduce a numerical method based on renewal processes to extract
the parameters of run-and-tumble dynamics from the experimental
measurements of the intermediate scattering function. }\fi We first validate our
approach against data produced using agent-based simulations. This
allows us to identify the length and time scales required for an
accurate measurement of the motility parameters, including
tumbling frequency and swim speed. We compare different models for the
run-and-tumble dynamics by accounting for speed variability at the
single-cell and population level, respectively. Finally, we apply our
approach to experimental data on wild-type \emph{Escherichia coli} obtained using differential dynamic microscopy. 
\end{abstract}
\maketitle

A hallmark of many living microorganisms is their ability to
self-propel in liquid environments~\cite{Romanczuk:2012,
  Elgeti:2015,Bechinger:2016}. To optimize their survival strategies,
such as foraging~\cite{benichou2011intermittent} and escaping from
harm~\cite{Wadhams:2004}, many microorganisms employ reorientation
mechanisms in addition to their directed swimming motion. During
reorientation events, cells change their swimming direction through
microorganism-specific processes. Examples range from the
`run-reverse(-flick)' patterns employed by several marine
bacteria~\cite{Taktikos:2013,Taute:2015} and
archaea~\cite{Thornton:2020}, to the `run-reverse-wrap' modes of
\emph{Pseudonmas putida}~\cite{Alirezaeizanjani:2020}, to the random
change of the swimming direction of the algae \emph{Chlamydomonas
reinhardtii} due to the buckling of one of their
cilia~\cite{Polin:2009}. The most widely studied microorganism is
probably the bacterium \emph{Escherichia coli}, which perform 
paradigmatic `run-and-tumble' (RT)
motion~\cite{Berg:1972,Schnitzer:1993,celani2010bacterial,chatterjee2011chemotaxis}.

The motion of \emph{E. coli} in bulk is characterized by two
alternating phases: a running and a tumbling phase. In the running
phase, bacteria swim persistently with an almost constant
speed. `Running' bacteria stochastically enter a tumbling phase due to
the random unbundling of one or several of their flagella. During a tumble, bacteria
undergo active rotational diffusion before they resume swimming
in a new direction \cite{saragosti:2012}. In a homogeneous environment, this RT motion
resembles a random walk at large length and time
scales~\cite{Berg:1972,Lovely:1975,Schnitzer:1993,Cates:2013}.  Such
RT motility pattern becomes drastically altered in the presence of surfaces,
which can induce circular motion via hydrodynamic
couplings~\cite{Lauga:2006, diLeonardo:2011}. Thus, the quantitative
characterization of the RT dynamics requires a 3D bulk study, far away
from perturbing boundaries.

Due to the significant biological importance of the tumbling statistics for, e.g., bacterial chemotaxis, the underlying genetics and biochemistry have been studied extensively~\cite{Wadhams:2004}. The dynamics of individual flagella have been measured \textit{in vitro} using single-motor experiments~\cite{cluzel2000ultrasensitive, korobkova2004molecular}. A quantitative characterization in 3D and at the population level, however, has remained a challenge. Tracking single-cell trajectories in 3D requires a low cell concentration and is limited to short  trajectories. The statistical accuracy is therefore often insufficient to reliably extract the tumbling statistics. 

Differential dynamic microscopy (DDM) is a possible alternative for
studying the motion of micro-swimmers~\cite{Cerbino:2008}. It measures
the intermediate scattering function (ISF), $f({\bf k},\tau)$, i.e. the
probability density of particle displacements $\Delta{\bf r}$ in Fourier space:
\begin{equation}
  f({\bf k},\tau) = \left\langle e^{-i {\bf k}\cdot \Delta{\bf r}(\tau)} \right\rangle\;.
\end{equation}
The ISF is usually computed by averaging over a large number of agents
($\gtrsim 10^4$ for \textit{E. coli} in Ref.~\cite{Wilson:2011}),
which provides reasonable statistics at the population level. Fitting
the experimentally measured ISF to a theoretical model can then give
access to the motility parameters of the microorganisms. While recent
work on catalytic Janus colloids has resolved the transition from
persistent motion to a randomization of the swimming direction due to
rotational diffusion in 2D~\cite{Kurzthaler:2018:janus}, the swimming properties of microorgansims
in bulk have only been measured up to length scales of a persistence
length~\cite{Wilson:2011,Martinez:2012,Jepson:2019,Croze:2019}.

Although quantifying the tumbling statistics at larger length scales using
DDM may appear to be merely an extension of previous
work~\cite{Wilson:2011}, it has remained out of reach so
far. Theoretically, the difficulty lies in the nonexistence of an
analytical form of the ISF in 3D in the time domain. Although its
Laplace transform is known~\cite{Angelani:2013,Martens:2012}, fitting
the ISF in Laplace space to the Laplace transform of the discrete data
is often problematic, due to the significant integration error
introduced by the finite data sequence. Thus, one has to calculate the
ISF in the time domain.

Here, we propose two strategies for calculating the ISF of RT
particles in real time, based on a renewal theory: (1) We solve the
renewal equations in Fourier space numerically, which immediately
provides the ISF, and (2) we compute the numerical inverse of the ISF
in Laplace space, which is known analytically. (We note that fitting the data in Laplace space always proves unpractical.) 
Both methods can, in
principle, be used equivalently and be applied for various
scenarios. However, in practice, they are suitable for different
situations: For particles with an intrinsic variability of swim speed
and with arbitrary running and tumbling time distributions, option (1)
is the only route. For particles with a speed variability at the
population level, the direct inverse Laplace transform (2) is easier
to compute, but this strategy is restricted to exponential running and
tumbling time distributions. 

The paper is organized as follows: In Sec.~\ref{sec:renewaltheory} we
introduce the renewal theory and compare the models with speed
variability at the single-cell and population level. In
Sec.~\ref{sec:parameters}, we present a robust protocol for the
parameter estimation, which we validate using simulated data. In
Sec.~\ref{sec:exp}, we analyze the data of wild-type \textit{E. coli}
using the model with speed variability at the population level and
compare it to the results from Ref.~\cite{Kurzthaler:2022}, where an
intrinsic variability of swimming speed was assumed. Finally we
summarize our results in Sec.~\ref{sec}.

\section{Renewal theory}
\label{sec:renewaltheory}

\begin{figure}[tp]
\centering
\includegraphics[width=\linewidth]{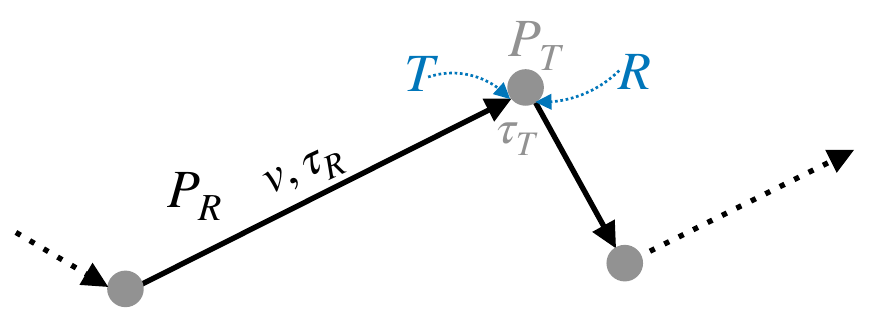}
\caption{Schematic of the run-and-tumble motion of a particle with mean run and tumbling times, $\tau_R$ and $\tau_T$, respectively. The particle moves at velocity $v$ during the run phase.
$P_R$ and $P_T$ are the probabilities of the swimmer to be in a run or tumbling phase. Further, $T$ and $R$ denote the probabilities to start tumbling or running, respectively.}
\label{fig:schematic}
\end{figure}

We consider a model of RT bacterium alternating between persistent runs
in quasi-straight lines and finite-duration tumbles during which the
cells fully randomize their
directions~\cite{Berg:1972,Schnitzer:1993,celani2010bacterial,chatterjee2011chemotaxis,Angelani:2013}.
The probability to find a bacterium displaced by a distance $\vec{r}$
after a lag time $\tau$ is $P(\vec{r},\tau)=P_R(\vec{r}, \tau)+
P_T(\vec{r}, \tau)$, where $P_R(\vec{r},\tau)$ and $P_T(\vec{r},\tau)$
correspond to the probability to be at position ${\bf r}$ after a lag
time $\tau$ and to be in a running or tumbling phase,
respectively. The ISF of non-interacting bacteria is then obtained via
a Fourier transform: $f_{RT}(\vec{k},\tau)= \int \!\diff^3 r
\exp(-\imath \, \vec{k}\cdot\vec{r})P(\vec{r},\tau)$. We denote by
$\varphi_{R}(\tau)$ and $\varphi_{T}(\tau)$ the distributions of the
durations of the RT phases, respectively. To allow for generic
distributions, which need not correspond to Markovian processes, we
also introduce the probabilities that a bacterium \textit{starts}
running or tumbling at displacement $\vec{r}$ and lag time $\tau$,
which we denote by $R(\vec{r},\tau)$ and $T(\vec{r},\tau)$,
respectively. Finally, the propagators $\mathbb{P}_R(\vec{r},\tau)$
and $\mathbb{P}_T(\vec{r},\tau)$ measure the probability that a
bacterium travels a distance $\vec{r}$ during a time $\tau$ in a
running or a tumbling phase, respectively.

To compute the ISF, we describe the RT dynamics as a renewal process~\cite{Feller:1971,Mendez:2014,Zaburdaev:2015} for which $P_R(\vec{r},\tau)$ satisfies 
\begin{equation}
P_R=P_R^0+\int_0^\tau\!\! \diff t\!\int\!\diff^3 \ell \  R(\vec{r}-\boldsymbol{\ell}, \tau-t)\varphi_R^0(t)\mathbb{P}_R(\boldsymbol\ell,t), \label{eq:prob_PR}
\end{equation}
where $\varphi^0_R(t)=\int_t^\infty\!\diff t' \ \varphi_R(t')$ is the probability that the run time exceeds $t$. Further, we denote the probability that the bacterium arrives in $\vec{r}$ at time $\tau$ without having tumbled in $[0,\tau]$ by $P^0_R(\vec{r},\tau) := p_R \mathbb{P}_R(\vec{r},\tau)\int_{\tau}^\infty \diff t \, \varphi_R(t)(t-\tau)/\tau_R$. The probability depends on the fraction of time the cell spends running, $p_R=\tau_R/(\tau_R+\tau_T)$, and on the average times spent running or tumbling, $\tau_{R,T}=\int_0^\infty\diff t \, t\varphi_{R,T}(t)$. Equation~\eqref{eq:prob_PR} states that the probability to be at $\vec{r}$ at time $\tau$ is the sum of the probabilities of arriving in $\vec{r}$ without tumbling in $[0,\tau]$, $P_R^0$, and with at least one tumble. In the second case, the last tumble takes place at arbitrary displacements $\vec{r}-\boldsymbol{\ell}$ and lag times $\tau-t$,  which should be summed over. Similarly, the probability that the bacterium starts a new run at displacement $\vec{r}$ and lag time $\tau$, $R(\vec{r},\tau)$, takes into account the possibility that this is the first run or that a run has already finished at $\tau-t$ and $\vec{r}-\boldsymbol\ell$:
\begin{equation}
R=R^1+\int_0^\tau\! \!\diff t\!\int\!\diff^3\ell \ T(\vec{r}-\boldsymbol\ell,\tau-t)\varphi_T(t)\mathbb{P}_T(\boldsymbol\ell,t)\;. \label{eq:prob_R}
\end{equation}
Here, $R^1(\vec{r},\tau):= (1-p_R)\mathbb{P}_T(\vec{r},\tau)\int_\tau^\infty \diff t \, \varphi_T(t)/\tau_T$ is the probability of starting the first run in $\vec{r}$ at time $\tau$. 
By swapping $R$ and $T$ everywhere in Eqs.~\eqref{eq:prob_PR}-\eqref{eq:prob_R} we obtain two other (formally identical) renewal equations for the probabilities $P_T(\vec{r},\tau)$ and $T(\vec{r},\tau)$:
\begin{align}
P_T&\!=\!P_T^0 +
\int_0^\tau\!\! \diff t\!\int\!\diff^3 \ell \  T(\vec{r}-\boldsymbol{\ell}, \tau-t)\varphi_T^0(t)\mathbb{P}_T(\boldsymbol\ell,t), \label{{eq:prob_PT}}\\
T&\!=\!T^1+\int_0^\tau \!\!\diff t\!\int\!\diff^3\ell \ R(\vec{r}-\boldsymbol\ell,\tau-t)\varphi_R(t)\mathbb{P}_R(\boldsymbol\ell,t). \label{eq:prob_T}
\end{align}

A RT dynamical model is then entirely determined by the propagators
$\mathbb{P}_{R,T}$ and by the choice of the reorientation process
specified by the distribution $\varphi_{R,T}$ from which $P_{R,T}^0$,
$R^1$, and $T^1$ follow. Since the renewal equations couple all
positions and times, they are hard to solve in
$\mathbf{r}$-space. Exploiting the convolution theorem, a Fourier
transform yields a set of equations that are decoupled in Fourier
space. Thanks to the isotropy of the system, they depend only on the
wave number $k=|\vec{k}|$ and can be solved for each $k$
separately. Eqs.~\eqref{eq:prob_PR}-\eqref{eq:prob_T} lead to:
\begin{align}
P_R(k,\tau)&\!=\!P_R^0(k,\tau)\!+\!\int_0^\tau\!\! \diff t \ R(k, \tau\!-\!t)\varphi_R^0(t)\mathbb{P}_R(k,t),\label{eq:renewalFT1}\\
R(k,\tau)&\!=\!R^1(k,\tau)\!+\!\int_0^\tau\! \! \diff t \ T(k,\tau\!-\!t)\varphi_T(t)\mathbb{P}_T(k,t),\label{eq:renewalFT2} \\
P_T(k,\tau)&\!=\! P^0_T(k,\tau)\!+\!\int_0^\tau\!\!\diff t \ T(k,\tau\!-\!t)\varphi^0_T(t)\mathbb{P}_T(k,t),\label{eq:renewalFT3}\\
T(k,\tau) &\!=\!T^1(k,\tau)\!+\!\int_0^\tau\!\!\diff t \ R(k,\tau\!-\!t)\varphi_R(t)\mathbb{P}_R(k,t).\label{eq:renewalFT4}
\end{align}
Once a given RT dynamical model is chosen, Eqs.~\eqref{eq:renewalFT1}-\eqref{eq:renewalFT4} permit numerical evaluation of the ISF for RT particles, $f_{RT}(k,\tau)=P_R(k,\tau)+P_T(k,\tau)$.

\begin{figure*}[htp]
\includegraphics[width = \linewidth]{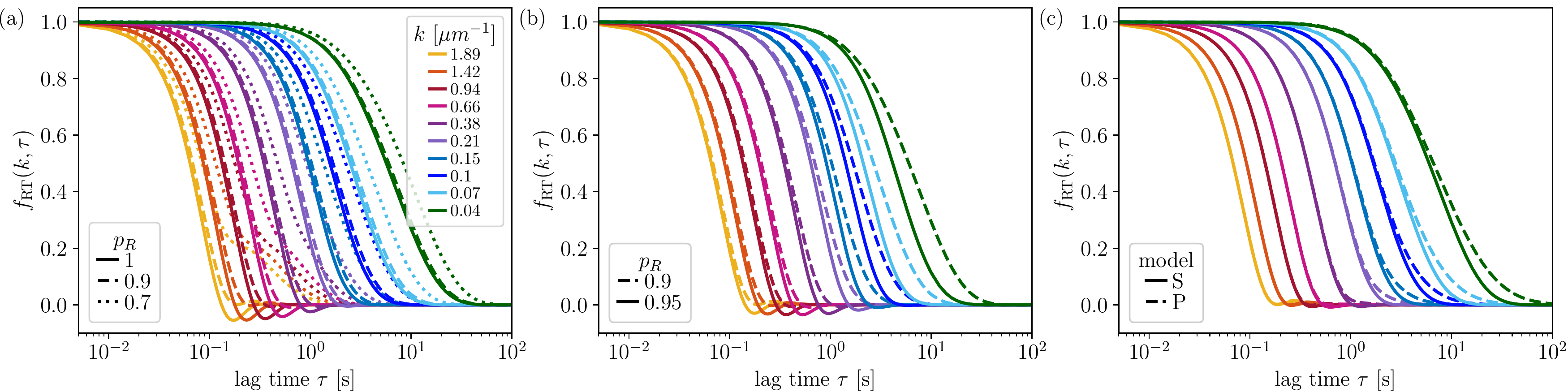}
\caption{ISFs, $f_{RT}(k,\tau)$, for our
  RT model.  (a-b) Intrinsic speed variability model for different
  fractions of run times $p_R=\tau_R/(\tau_R+\tau_T)$.  In (a) we vary
  the tumbling time with parameter values $\tau_R=1$~s and $\tau_T =
  0,0.1,0.5~\text{s}$ and in (b) we vary the run time, i.e.,
  $\tau_T=0.1$~s and $\tau_R = 2,1$~s. The other motility parameters
  are $\bar v = 15\, \upmu\text{m}\text{s}^{-1}$, $\sigma_v=4.5\,
  \upmu\text{m}\text{s}^{-1}$, and $D=0.3\,
  \upmu\text{m}^2\text{s}^{-1}$.  (c) Comparison of the ISF for the
  model with speed fluctuations at the single cell ($S$; solid line)
  and population ($P$; dashed line) level using identical parameters ($\tau_R=1$~s, $\tau_T = 0.1~\text{s}$, $\bar v = 15\, \upmu\text{m}\text{s}^{-1}$, $\sigma_v=4.5\,
  \upmu\text{m}\text{s}^{-1}$, and $D=0.3\,
  \upmu\text{m}^2\text{s}^{-1}$).
\label{fig:ISF_suppl}}
\end{figure*}

Inspection of Eqs.~\eqref{eq:renewalFT1}-\eqref{eq:renewalFT4} suggests that analytical progress can be made in Laplace space, following~\cite{Angelani:2013}. In particular, a  Laplace transform of Eqs.~\eqref{eq:renewalFT1}-\eqref{eq:renewalFT4} yields the propagators, $P_R$ and $P_T$, for arbitrary RT distributions, $\varphi_R$ and $\varphi_T$, and probabilities, $\mathbb{P}_R$ and $\mathbb{P}_T$:

\begin{widetext}
\begin{align}
P_R(k,s) &= P_R^0(k,s)+\mathcal{L}\left[\varphi_R^0(\tau)\mathbb{P}_R(k,\tau)\right](s) \frac{R^1(k,s)+T^1(k,s)\mathcal{L}\left[\varphi_T(\tau) \mathbb{P}_T(k,\tau)\right](s)}{1- \mathcal{L}\left[\varphi_R(\tau) \mathbb{P}_R(k,\tau)\right](s)\mathcal{L}\left[\varphi_T(\tau) \mathbb{P}_T(k,\tau)\right](s)}, \label{eq:PR_Laplace_time}\\
P_T(k,s) &= P_T^0(k,s)+\mathcal{L}\left[\varphi_T^0(\tau)\mathbb{P}_T(k,\tau)\right](s) \frac{T^1(k,s)+R^1(k,s)\mathcal{L}\left[\varphi_R(\tau) \mathbb{P}_R(k,\tau)\right](s)}{1- \mathcal{L}\left[\varphi_T(\tau) \mathbb{P}_T(k,\tau)\right](s)\mathcal{L}\left[\varphi_R(\tau) \mathbb{P}_R(k,\tau)\right](s)}, \label{eq:PT_Laplace_time}
\end{align}
\end{widetext}
where $\mathcal{L}[f(\tau)](s):= \int_0^\infty\diff \tau \exp(-s\tau)f(\tau)$ denotes the Laplace transform of a function $f(\tau)$.

\subsection{Intrinsic speed variability}

So far we have introduced the general framework of a renewal process
switching between the running and tumbling phases. Let us now specify
the time distributions $\varphi_{R,T}$ and the propagators
$\mathbb{P}_{R,T}$ for RT particles. We first consider the case in
which a bacterium changes its swim speed after each tumble, which we
refer to as \emph{intrinsic speed variability}. This accounts for
the fluctuations of the propulsion speed over time that have recently
been reported experimentally~\cite{Turner:2016}. Alternatively, the
distribution of swimming speed can be accounted for at the population
level, leading to a different model discussed in
Sec.~\ref{sec:population_variability}.

For simplicity, we here consider exponential distributions for the run
and tumbling times with $\varphi_{R,T}(t)=
\exp(-t/\tau_{R,T})/\tau_{R,T}$, where $\tau_R$ and $\tau_T$ denote
the mean run and tumbling durations, respectively. We note, however,
that our formalism allows discussing more general distributions as
well. We now discuss the propagators $\mathbb{P}_R$ and
$\mathbb{P}_T$.

Assuming that tumbling particles diffuse with diffusivity $D$, the corresponding propagator is given by
\begin{equation}\label{eqn:PT_intrinsic}
\mathbb{P}_T(k,\tau)=\exp(-Dk^2\tau).
\end{equation}
For a swimming particle with speed $v$ and thermal diffusion constant
$D$, the propagator instead reads
$\exp(-Dk^2\tau)\sin(vk\tau)/(vk\tau)$. Assuming that, after each tumble,
the particle samples a new swimming speed from a distribution $p(v)$,
the propagator of the swimming particles is
\begin{equation}
\mathbb{P}_R(k,\tau)=\int_0^\infty p(v)\exp(-Dk^2\tau)\frac{\sin(vk\tau)}{vk\tau}\diff v.
\end{equation}
We use the Schulz distribution which is characterized by a mean velocity $\bar v$ and standard deviation $\sigma_v$~\cite{Martinez:2012},
\begin{equation}
p(v)=\frac{v^Z}{\Gamma(Z+1)}\left(\frac{Z+1}{\bar v}\right)^{Z+1}e^{-(Z+1)v/\bar v}, \label{eqn:schultz}
\end{equation}
with $Z=\bar v^2/\sigma_v^2-1$. Then $\mathbb{P}_R(k,\tau)$ can be computed analytically as
\begin{equation}\label{eqn:PR_intrinsic}
\mathbb{P}_R(k,\tau)=e^{-Dk^2\tau}\left(\frac{Z+1}{Zk\bar v\tau}\right)\frac{\sin(Z\arctan\xi)}{(1+\xi^2)^{Z/2}},
\end{equation}
with $\xi=k\bar v\tau/(Z+1)$.

Using Eqs.~\eqref{eqn:PT_intrinsic} and \eqref{eqn:PR_intrinsic} as
input, we can solve the integral
equations~\eqref{eq:renewalFT1}-\eqref{eq:renewalFT4} numerically by
time stepping $\tau$ for each wave number separately. This then leads
to the intermediate scattering function (ISF) for the
run-and-tumble particle, $f_{RT}(k,t)=P_R(k,\tau)+P_T(k,\tau)$.

We note that the Laplace transform obtained from
Eq.~\eqref{eq:PR_Laplace_time} and \eqref{eq:PT_Laplace_time} involves
hypergeometric functions which makes its numerical inversion
cumbersome and inefficient using Weeks'
method~\cite{Weeks:1966,Weideman:1999}.

\subsection{Speed variability at the population level}\label{sec:population_variability}

Alternatively, one could consider a model, where the speed $v$ of a
given bacterium is constant, but where $v$ is distributed over the
bacterial population. As we show below, this leads to a simpler
expression in Laplace space, but does not account for temporal
fluctuations of $v$ at the single-bacterium level. Again, the speed
distribution $p(v)$ of the population is chosen to be a Schultz
distribution. In particular, within the renewal framework we replace
the propagator for the run phase by $\mathbb{P}_R(k,\tau)=\exp(-Dk
^2\tau) \sin(vk\tau)/(vk\tau)$. Then the ISF is obtained by
post-averaging the ISF of a single cell over the speed distribution,
$f_{RT}(k,\tau)=\int \diff v\,p(v)[P_R(k,\tau)+P_T(k,\tau)]$. The two
models are not equivalent due to the non-linearity of
Eqs.~\eqref{eq:PR_Laplace_time}-\eqref{eq:PT_Laplace_time} with
respect to the propagators $\mathbb{P}_{R,T}$.

For this model, the numerical evaluation of the ISF, $f_{RT}(k,\tau)$,
is expensive within the renewal framework due to the final integration
over $p(v)$. Fortunately, the Laplace transform, $f_{RT}(k,s)$ is
simpler than in the intrinsic-speed-variability model. Substituting
the propagators, $\mathbb{P}_R(k,\tau)$ and
$\mathbb{P}_T(k,\tau)=\exp(-Dk ^2\tau)$, and the exponential RT
distributions, $\varphi_R$ and $\varphi_T$, into
Eqs.~\eqref{eq:PR_Laplace_time} and \eqref{eq:PT_Laplace_time}, the
ISF in Laplace space for a population of non-interacting RT particles
with speed distribution reads:
\begin{widetext}
\begin{equation}
f_{RT}(k,s)=\int_0^\infty \diff v\,p(v)\frac{kv\tau_T^2\tau_R+\tau_R(\tau_R+2\tau_T+\tau_T\tau_R(Dk^2+s))\arctan(kv\tau_R/(Dk^2\tau_R+1+s\tau_R))}{(\tau_R+\tau_T)[kv\tau_R(1+\tau_T Dk^2+\tau_Ts)- \arctan(kv\tau_R/(Dk^2\tau_R+1+s\tau_R))]}\;. \label{eq:ISF_RTP_const_speed}
\end{equation}
\end{widetext}
Note that for exponentially distributed RT times, the ISF in Laplace
space could also be obtained by generalizing the method introduced in
Ref.~\cite{Martens:2012} for finite-duration tumbles. The time-domain
ISF $f_{RT}(k,\tau)$ can be computed from
Eq.~\eqref{eq:ISF_RTP_const_speed} using the standard Weeks'
method~\cite{Weeks:1966,Weideman:1999}. Before discussing our fitting
procedure and its validation using simulated data in
Sec.~\ref{sec:parameters}, we first show below how the ISFs depend on
the ingredients of the RT dynamics and the source of speed
fluctuations.

\subsection{Intermediate scattering functions}
Fig.~\ref{fig:ISF_suppl}(a-b) show the ISFs for the RT model with intrinsic speed variability evaluated for motility parameters measured for \emph{E. coli} (see figure caption). 
The calculated ISFs $f_{RT}(k,\tau)$ [Fig.~\ref{fig:ISF_suppl}(a)] show a clear evolution at short times and small length scales $\ell=2\pi/k$ as one varies the RT durations close to their estimated biological values of $\tau_T\simeq \SI{0.1}{\second}$ and $\tau_R\simeq \SI{1}{\second}$~\cite{Berg:1972}. For instantaneous tumbles, $\tau_T= 0$ ($p_R=1$), the ISFs display oscillations for large $k$, which are smeared and disappear at times $\tau\gtrsim\tau_R$ and at small wave numbers $k\lesssim \SI{0.38}{\per\micro\meter}$ corresponding to a length scale $\ell\approx \SI{16.5}{\micro\meter}$, comparable with the persistence length $\ell_p=\langle v \rangle\tau_R=\SI{15}{\micro\meter}$ beyond which the motion becomes randomized by tumbles. As the tumble duration increases ($p_R$ decreases), oscillations fade out  until a hump develops ($\tau_T=\SI{0.5}{\second}$, $p_R\approx0.7$) due to the diffusive motion of tumbling bacteria at small~$\ell$. 

Fig.~\ref{fig:ISF_suppl}(b) shows the ISFs for a fixed tumble duration of $\tau_T=0.1$~s and varying run time $\tau_R=2,1$~s. As the run time increases, we observe stronger oscillations at short times $\tau\lesssim\tau_R$ and large wave numbers $k$ and a more rapidly decreasing ISF at small wave numbers, which indicates that the regime of effective diffusion emerges at larger length and time scales.
This pattern of behavior suggests that, in principle, experimentally-measured ISFs should contain enough information to characterize all features of the RT dynamics of \Ecoli, including the tumbling statistics. 

\begin{figure*}
\centering
\includegraphics[width = \linewidth, keepaspectratio]{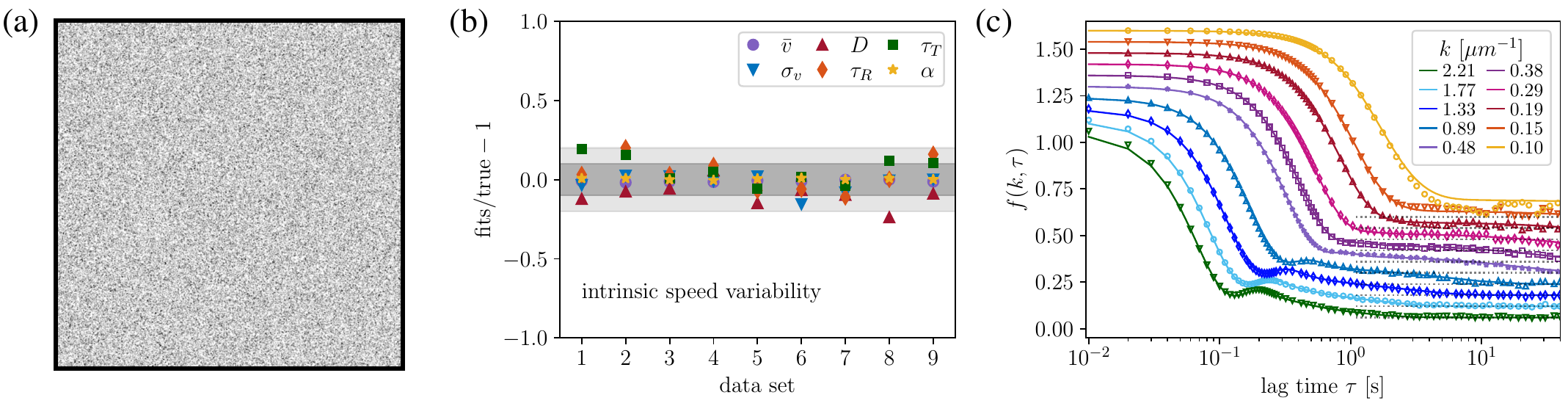}
\caption{(a) Snapshot of a RT simulation with
  $1\times$~magnification. (b-c) Validation of theoretical predictions
  with simulations for nine parameter sets with $\bar v = 17\,
  \upmu\text{ms}^{-1}$, $\sigma_v = 4.3\, \upmu\text{ms}^{-1}$,
  $D=0.3\, \upmu\text{m}^2\text{s}^{-1}$, $\alpha = 0.9$,  varying the
  mean run and tumbling times, $\tau_R \in [0.5,1.5] \ \text{s}$ and
  $\tau_T \in [0.1,0.5] \ \text{s}$.  (b) Parameter estimates obtained
  by fitting theoretical predictions of the ISF to agent-based
  simulations.  The estimates are compared with the true parameters
  for the nine data sets. They were extracted from a global fit
  including wave numbers $k\in[0.15,2.21] \ \upmu\text{m}^{-1}$. Dark
  gray regions correspond to fitted parameters within $\pm10\%$
  of the true values (light gray corresponds to $\pm20\%$).  (c)
  ISF, $f(k,\tau)$, for $\tau_R=1.25$~s, $\tau_T=0.2$~s (data set 8)
  and different wave numbers $k$. Theoretical predictions and
  simulations correspond to lines and symbols, respectively. The
  ISFs are shifted w.r.t. the $y-$axis and the gray dotted lines
  indicate $y=0$.  }
\label{fig:validation}
\end{figure*}

Finally, we compare the ISFs for both models, see
Fig.~\ref{fig:ISF_suppl}(c). In particular, we observe that at short
times and length scales the ISFs of both models are almost
indistinguishable and hence the fingerprint of the speed variability
on the ISFs is subtle. The effect, however, becomes visible in our
theory at large length scales, corresponding to $k\gtrsim
\SI{0.38}{\per\micro\meter}$. Whether this small difference of the
ISFs will be measurable and identifiable from experiments, will be
discussed later.

\section{Numerical protocol and validation on simulated data}
\label{sec:parameters}
We next present the numerical protocol that allows us to estimate the
motility parameters of bacteria, such as their mean RT times and swimming
speed, from measured ISFs. For the sake of completeness, we first recall
below how ISFs are measured experimentally using differential dynamic microscopy (DDM).

\subsection{Differential dynamic microscopy}\label{sec:ddm}
DDM is a high-throughput method that
provides quantitative information on 3D swimming microorganisms
through their ISF, see Refs.~\cite{Cerbino:2008,
  Martinez:2012} for details.  Briefly, the differential image
correlation function (DICF), $g(\vec{k},\tau)$, i.e., the power
spectrum of the difference between pairs of images separated by time
$\tau$, is obtained via $g(\vec{k}, \tau)=\left\langle\left|I(\vec{k},
t+\tau)-I(\vec{k}, t)\right|^2\right\rangle_t$, where $I(\vec{k},t)$
is the Fourier transform of the image $I(\vec{r},t)$ and
$\langle\cdot\rangle_t$ denotes an average over time $t$. Under
suitable imaging conditions and for isotropic motion, the DICF is
related to the ISF \cite{Wilson:2011,Martinez:2012,Reufer2012},
$f(k,\tau)$, via
\begin{align}
g(k,\tau) =\left \langle g(\vec{k},\tau)\right\rangle_{\vec{k}}= A(k)\left[1-f(k,\tau)\right]+B(k)
\end{align}
with $\left\langle\cdot\right\rangle_\vec{k}$ denoting average over $\vec{k}$ and $A(k)$ and $B(k)$ the signal amplitude and instrumental noise, respectively. These coefficients are obtained from the plateau of $g(k,\tau)$ at long and short times, where the ISF approaches $f(k,\tau\to \infty)\rightarrow 0$ and $f(k,\tau\to 0)\rightarrow 1$, respectively.

\subsection{Fitting procedure}\label{sec:fitting}
To reliably extract quantitative information from the measured ISFs using our renewal theory, we implement a fitting procedure based on the minimization of the squared errors using a Nelder Mead optimization algorithm~\cite{Nelder:1965}. We apply a multi-start fitting analysis, where several fits are obtained for various initial values and the parameter set yielding the smallest error is chosen. In most fitting procedures several initial values provided the same result, which strengthens the reliability of our procedure.

We performed a global fit including data for several wave numbers simultaneously. Using one dataset, we tested several wave number ranges and found the most adequate should include length scales of the order of the cell body, up to length scales resolving the randomization of the swimming direction, i.e.  $k\ell \lesssim 2\pi$. The parameter estimation method has been validated with simulations (see Section~\ref{sec:validation}).

\subsection{Validation of the fitting procedure}\label{sec:validation}

Before tackling the experimental data, we have validated our parameter estimation method with computer simulations (see Appendix~\ref{appendix:sim}). To do so, we consider exponentially distributed RT times with different $\tau_R$ and $\tau_T$, which are close to those reported previously~\cite{Berg:1972}. In particular, we set $\tau_R\in[0.5,1.5]\,$s and $\tau_T\in[0.1, 0.5]\,$s corresponding to $p_R = \tau_R/(\tau_R+\tau_T) \in [0.67,0.91]$. 
We further choose values for the remaining motility parameters, including the mean velocity, $\bar v$, its standard deviation, $\sigma_v$, the translational diffusivity, $D$, that are typical for \Ecoli\ suspensions~\cite{Wilson:2011,Martinez:2012}. Following experimental findings \cite{Kurzthaler:2022}, we add a fraction $1-\alpha$ of non-motile cells that undergo Brownian motion with diffusivity $D$ in the simulations. Thus the ISF obtained from simulated data should follow
\begin{equation}
    f(k,\tau)=\alpha f_{RT}(k,\tau)+(1-\alpha)e^{-Dk^2\tau}\;.
\end{equation}

First, we perform simulations of the intrinsic-speed variability
model. We employ a global fitting procedure (as outlined in
Section~\ref{sec:fitting}) and simultaneously include wave numbers
$k\in[0.15,2.21]\,\upmu$m$^{-1}$.  Fitting our renewal theory to the
ISF extracted from simulated data of particles with intrinsic speed
variability reveals that our fitting protocol reliably reproduces the
true parameter values. In particular, most of the fitted parameters
lie within $\pm10\%$ error with respect to the true values
[Fig.~\ref{fig:validation}(b)].  Figure~\ref{fig:validation}(c) shows
excellent agreement between the simulated ISF with mean run and
tumbling times, $\tau_R=1.25\,$s, $\tau_T=0.2\,$s (dataset 8), and the
theoretical predictions obtained from the fitting procedure.

\begin{figure*}
    \centering
    \includegraphics[width = \linewidth, keepaspectratio]{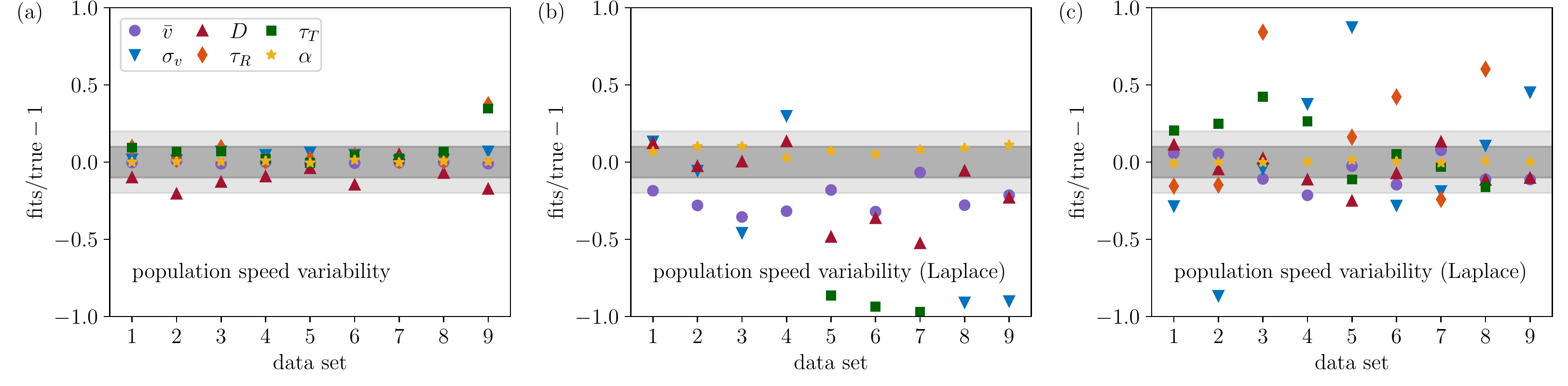}
    \caption{Parameter estimates obtained by fitting theoretical
      predictions of the ISF to agent-based simulations with speed
      variability at the population level. Parameters are the same as
      in Fig.~\ref{fig:validation}: $\bar v = 17\,
      \upmu\text{ms}^{-1}$, $\sigma_v = 4.3\, \upmu\text{ms}^{-1}$,
      $D=0.3\, \upmu\text{m}^2\text{s}^{-1}$, $\alpha = 0.9$. The mean
      run and tumbling times are varied within $\tau_R \in [0.5,1.5]
      \ \text{s}$ and $\tau_T \in [0.1,0.5] \ \text{s}$.  (a) Fitting
      in lag time $\tau$ with numerical inverse Laplace transform of
      theoretical ISF. (b-c) Fitting in Laplace time $s$ using the
      numerical Laplace transform of the data, with $s$ ranging from
      (b) $1/\tau_{\rm max}$ to $1/\tau_{\rm min}$ and (c)
      $10/\tau_{\rm max}$ to $1/\tau_{\rm min}$.  The estimates are
      compared with their true values for the nine data sets. Outliers
      with an error such that (fits/true$-1)>1$ are not shown. We
      used a global fit including wave numbers
      $k\in[0.15,2.21] \ \upmu\text{m}^{-1}$. Dark gray regions correspond to fitted parameters within $\pm10\%$ error of the true
      values (light gray corresponds to $\pm20\%$). }
    \label{fig:population}
\end{figure*}

We also perform simulations of a mixture of particles with different
speeds, fixing each particle speed during the simulation. We use the
inverse Laplace transform of Eq.~\eqref{eq:ISF_RTP_const_speed} to
calculate the ISF. Including a fraction $\alpha$ of
non-motile cells, we follow the same global fitting procedure as
introduced in Section~\ref{sec:fitting} to fit the data.  We find that
the fitted parameters again lie within $\pm10\%$ error with respect to the
true values [Fig.~\ref{fig:population}(a)].

Note that fitting the
expression~\eqref{eq:ISF_RTP_const_speed} directly in Fourier-Laplace
space over the full range $s\in[1/\tau_\text{max}, 1/\tau_\text{min}]$
leads to large, systematic errors for most parameters
[Fig.~\ref{fig:population}(b)]. These errors are mainly due to the loss
of numerical precision during integration of the discrete data to
calculate their Laplace transform. Using a smaller range
$s\in[10/\tau_\text{max}, 1/\tau_\text{min}]$
[Fig.~\ref{fig:population}(c)] slightly improves the fitting results
but does not lead to satisfactory estimates, especially for the
tumbling and running durations. We further note that while an optimized
$s-$range may lead to a satisfactory fit in Laplace space, it requires
 \textit{a priori} knowledge of the swimming parameters that makes
this procedure unsatisfactory. This is a major issue for the analysis
of experimental data, which probably explains why DDM has not been
used so far to characterize RT dynamics, despite the explicit
expressions for the propagators in Fourier/Laplace
space~\cite{Angelani:2013,Martens:2012}. For numerics as well as
experimental data, we found that the renewal theory was always a more
efficient and reliable avenue.

\section{Experiments}\label{sec:exp}
We now demonstrate that our numerical protocol can indeed be used to
characterize quantitatively experimental data. To do so, we use the
data on the wild-type AB1175 \textit{E. coli} strain presented in our
joint work~\cite{Kurzthaler:2022}. {In short, these data were obtained by 
imaging swimming cells in sealed capillaries on a fully-automated inverted bright-field microscope with a sCMOS camera. To characterize the RT dynamics we require access to length scales larger than the cells' persistence length, $\gtrsim \ell_p$, in 3D. Therefore, a large depth of field at low $k$ is needed to ensure that bacteria remain in view over large distances in all directions. To measure the dynamics at all relevant length scales, we consecutively recorded movies at $2\times$ and $10\times$ magnifications to extract the ISF for $k<0.9\, \upmu$m$^{-1}$ and $k\geq 0.9\, \upmu$m$^{-1}$ respectively using standard DDM procedures \cite{Wilson:2011,Martinez:2012}. We refer to the Supplemental Material of Ref.~\cite{Kurzthaler:2022} for more details on the experimental procedure. Fitting the ISFs to our renewal theory using the numerical protocol described here yields RT motility parameters. }

\begin{figure}[tp]
\centering
\includegraphics[width=\linewidth]{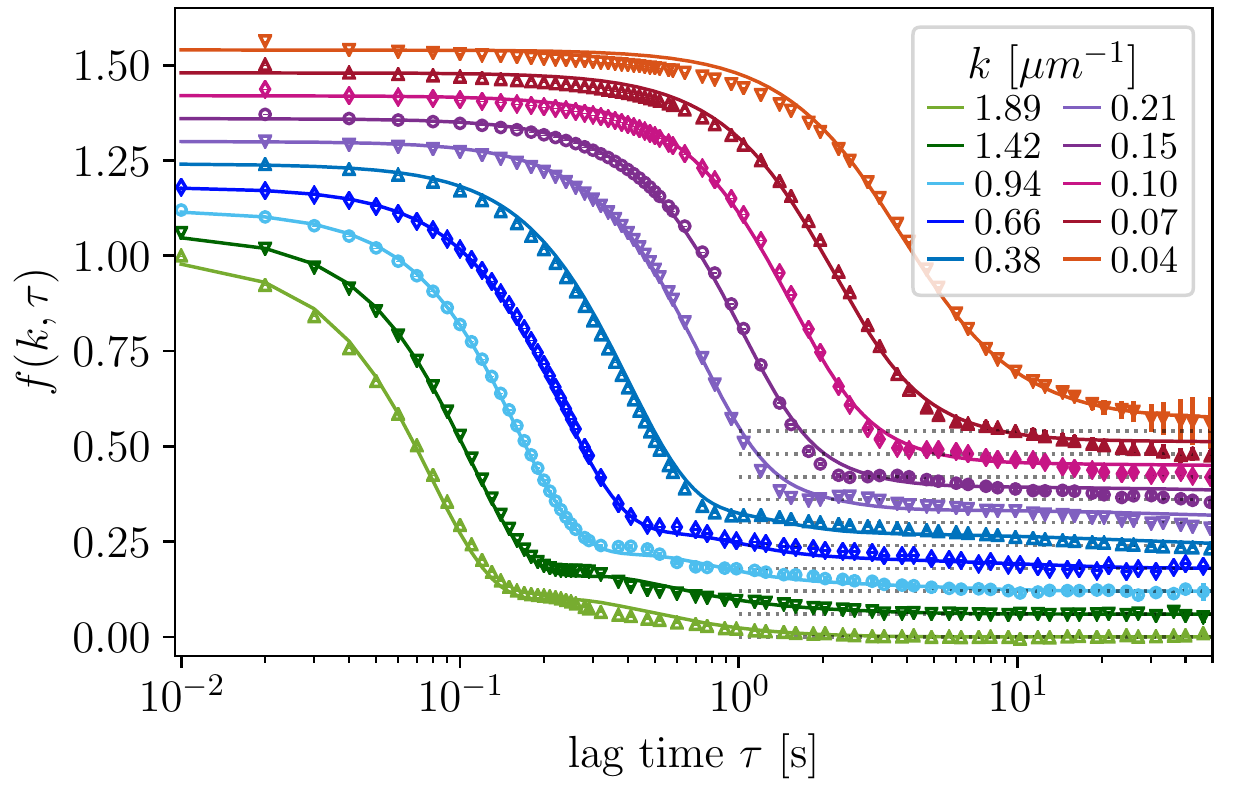}
\caption{ISFs for different wave numbers $k$. Symbols represent
  experimental results for \emph{E. coli} bacteria (same as in Fig.~3
  of Ref. \cite{Kurzthaler:2022}) and lines are the theoretical
  predictions obtained by considering speed variability at population
  level. The data has been fitted to the numerical inverse of
  Eq.~\eqref{eq:ISF_RTP_const_speed} including wave numbers $k\in [0.04, 1.89]\SI{}{\micro\meter}^{-1}$.
\label{fig:ISF_laplace}}
\end{figure}

We have reported the fitting
results from the intrinsic-speed-variability model in Fig.~4 of
Ref.~\cite{Kurzthaler:2022}. Here we fit the same data to the
theoretical predictions of the model that incorporate speed
fluctuations at the bacterial population level, using the numerical inverse
Laplace transform of Eq.~\eqref{eq:ISF_RTP_const_speed}. We obtain the
following motility parameters: 
$\alpha=97\pm0.3\%$, $\bar v=16\pm0.2\,\si{\micro\meter\per\second}$,
$\sigma_v=5.80\pm0.29\,\si{\micro\meter\per\second}$,
$D=0.25\pm0.04\,\si{\square\micro\meter\per\second}$,
$\tau_R=3.21\pm0.38\,\text{s}$, $\tau_T=0.50\pm0.05\,\text{s}$.  The
corresponding ISFs are shown in Fig.~\ref{fig:ISF_laplace} and display
nice agreement with the experimental data.

The estimates for the motility parameters are largely consistent with
those reported in Ref.~\cite{Kurzthaler:2022}, although they were obtained from fitting a slightly different model. We note the fraction of running time
$\tau_R/(\tau_R+\tau_T)= 0.866$ agrees with the well-cited results
from Berg and Brown~\cite{Berg:1972} and the fits in
Ref. \cite{Kurzthaler:2022} ($\tau_R/(\tau_R+\tau_T)=0.863$). This
suggests that the origin of the speed variability is indistinguishable
under the spatio-temporal scales measured in the experiments.  The run
and tumbling times obtained from the model with speed variability at
the population level are both slightly larger than that from the
intrinsic-speed-variability model, with an uncertainty of $\sim 10\%$
which is also slightly larger than that of the
intrinsic-speed-variability model ($\sim 5\%$ \cite{Kurzthaler:2022}).
We note that in real biological systems, both types of fluctuations
are expected; they do not lead to major differences as far as
displacement statistics are concerned. We note that implementing the
speed variability at the population level allows to work with explicit
expression for $f(k,s)$, which, however, have to be inverted
numerically.

\section{Summary and conclusion}
\label{sec}
In this paper, we developed a numerical protocol to quantitatively
characterize the tumbling statistics of run-and-tumble bacteria from
DDM measurements. First, we showed how to use the renewal theory to
calculate the intermediate scattering function of run-and-tumble
particles. We proposed two slightly different models, which account
for the speed fluctuations at either the intrinsic or the population
level. Then we demonstrated a robust protocol to extract parameters
from experimental data. The protocol was validated using agent-based
simulations and then applied to the experimental data of a wild-type
\emph{E. coli} AB1157 strain, which was reported in an accompanying paper~\cite{Kurzthaler:2022}. At the spatio-temporal scales of the
experiment, the two models seem to be indistinguishable and the
bacteria may in fact exhibit both: intrinsic speed variations and a
speed variability at the population level. In~\cite{Kurzthaler:2022},
we also show how our method can be employed to characterize a
transition between perpetual tumbling and smooth swimming in an
engineered bacterial strain whose tumbling statistics is under the
(now-quantitative) control of a chemical inducer.

The framework of renewal processes is not limited to the RT motion of bacteria and can be extended to other multi-mode motility patterns, such as the `run-reverse(-flick)'~\cite{Taktikos:2013,Taute:2015, Thornton:2020} or `run-reverse-wrap'~\cite{Alirezaeizanjani:2020} motion. In future work, our method may thus allow for a quantitative characterization of a large variety of microorganisms.
Furthermore, the numerical protocol proposed in this paper has vast potential applications, ranging from quantitatively studying the tactic response of individual cells to investigating complex collective bacterial organizations.

In particular, to gain a complete picture of bacterial chemotaxis on the cell level, the regulation of the tumbling statistics due to the presence of spatially varying, external chemical fields~\cite{clark-pnas-2005, kafri-prl-2008} may be established experimentally by using spatially-resolved DDM \cite{Reufer2012}. The `run-and-tumble' motion has been proposed as a paradigmatic model not only for \emph{E. coli}, but also for many other microorganisms, such as \emph{Euglena gracilis}~\cite{tsang2018polygonal}. The latter can direct its motion in the presence of light sources and its phototaxis has rich features because the cell can sense both the intensity and the polarization of light \cite{yang:2021}. Our framework may allow for a quantitative description of this intricate behavior and the associated tumbling statistics.

Finally, the regulation of the tumbling statistics of engineered
bacterial strains~\cite{Kurzthaler:2022,mckay:2017} may be exploited
for creating complex
self-organizations~\cite{Liu:2011,Curatolo:2020}. There our
high-throughput method could both help validate the design of the
engineered strains as well as allow for their quantitative
characterization.

\begin{acknowledgments}
This work was supported by the Austrian Science Fund (FWF) via
P35580-N and the Erwin Schr{\"o}dinger fellowship (J4321-N27), the
European Research Council Grant AdG-340877-PHYSAPS, the ANR grant
Bactterns, the Shenzhen Peacock Team Project (KQTD2015033117210153),
and the National Key Research and Development Program of China
(2021YFA0910700).
\end{acknowledgments}

\section*{Appendix}
\appendix
\section{Jackknife resampling method}
We obtain an estimate of the standard deviation $\text{SE}(\hat{p})$
of the fitted parameters $\hat{p}\in\{\langle v \rangle, \sigma_v,
\tau_R,\tau_T,D,\alpha\}$ using the Jackknife resampling
method~\cite{Efron:1981}. In our fitting procedure we fit datasets
corresponding to $N$ wave numbers $k=\left\{k_1,\dots,k_N\right\}$
simultaneously. The standard Jackknife resampling method is based on
estimating parameters $p_i$ by omitting the dataset corresponding to
wave number $k_i$ and thus fitting datasets corresponding to
$k=\left\{k_1,\dots k_{i-1},k_{i+1},\dots,k_N\right\}$ only. Repeating
this procedure for all wave numbers permits to estimate the Jackknife
standard deviation as
\begin{align}
\text{SD}(\hat{p}) &= \left[\frac{N-1}{N} \sum_{i=1}^N \left(\bar{p}-p_i\right)^2\right]^{1/2},
\end{align}
where $\bar{p}=\sum_i p_i/N$ denotes the average over all parameters $p_i$.
The errors of the fitted parameters are then obtained by $\text{SD}(\hat{p})$.

\section{Agent-based simulations}
 \label{appendix:sim}
We have performed agent-based simulations of non-interacting particles
in continuous space and time to validate our theoretical predictions
and parameter estimation
method following previously developed protocols~\cite{Wilson:2011,Martinez:2012}. First, we define two types of
particles: active agents, which perform run-and-tumble motion, and
passive particles, which only diffuse. The active particles alternate
between run and tumbling phases. The particles in a run phase move
along their swimming direction at constant speed and undergo
translational Brownian motion with diffusivity $D$. The Langevin
equation for the $i$-th running particle reads
\begin{align}
\frac{\diff\bm{r}_i}{\diff t}&=v_i\bm{u}_i+\sqrt{2D}\bm{\xi}_i\; , \label{langevin1}
\end{align}
where the components $\xi_{i,\alpha}$ are uncorrelated Gaussian white noise such
that $\langle\xi_{i,\alpha}(t)\xi_{j,\beta}(t')\rangle=\delta_{ij}\delta_{\alpha\beta}\delta(t-t')$ and $\vec{u}_i$ is the swimming direction of the $i$-th particle with unit length, $|\vec{u}_i|=1$. The swimming speed $v_i$ of the $i$-th particle after each tumbling event is sampled according to the Schultz distribution in Eq.~\eqref{eqn:schultz}. We note that accounting for a speed variability at the population level corresponds to keeping $v_i$ fixed for each cell throughout the simulation, but attributing different $v_i$ to different active particles according to Eq.~\eqref{eqn:schultz}.
The tumbling and passive particles perform only translational diffusion with corresponding Langevin equation for the $j$-th tumbling/passive particle,
$\diff\bm{r}_j/\diff t=\sqrt{2D}\bm{\xi}_j$,
where $D$ corresponds to the same diffusion coefficient as for running particles. The run time $\tau_i$ is distributed according to an exponential distribution with mean run duration $\tau_R$, $\varphi_R(\tau_i)=e^{-\tau_i/\tau_R}/\tau_R$, and the tumbling time $\tau'_j$ is distributed according to $\varphi_T(\tau'_j) = e^{-\tau'_j/\tau_T}/\tau_T$, with mean tumbling duration $\tau_T$. The mean run  and tumbling durations, $\tau_R$ and $\tau_T$, are the same for all  particles.

We randomly suspend $N \alpha \tau_R/(\tau_R+\tau_T)$ running
particles, $N \alpha \tau_T/(\tau_R+\tau_T)$ tumbling particles, and
$(1-\alpha)N$ passive particles in a cuboid container of size $L\times
L\times H$ with periodic boundary conditions. We sample the switching
time between run and tumbling phases using a Gillespie stochastic
algorithm for all the particles. Then by keeping track of the run and
tumbling states of each individual particle, we update the positions
of the particles according to the Langevin equations.  We generate
snapshots at time interval $\Delta t$ by taking a region
$-l/2<x,y<l/2,\ -h/2<z<h/2$ in the container, where $l<L$ is the size
of the image and $h<H$ is the depth of field, and by selecting all
particles suspended in this volume. The snapshot is an image of
$N_p\times N_p$ pixels, with pixel size $\delta l=l/N_p$. Then the
particle at position $(x,y,z)$ appears in the pixel $(n_x,
n_y)=(\lfloor x/\delta l\rfloor, \lfloor y/\delta l\rfloor)$ and its
eight neighborhoods, if any of the nine pixels are in the range
$[0,N_p-1]\times[0,N_p-1]$. Denoting $\delta x=x/\delta l-n_x$,
$\delta x'=\min\{\delta x, 1-\delta x\}$, $\delta y=y/\delta l-n_y$,
and $\delta y'=\min\{\delta y, 1-\delta y\}$, the intensity of a
single particle is spread out over the nine adjacent pixels as
\begin{equation}
I_{n_x+m,n_y+l}=C(z)I_m(\delta x)I_l(\delta y)\; ,
\end{equation}
where $m,l=0,\pm 1$, $C(z)=1-4z^2/h^2$ is the contrast function, and
\begin{subequations}
\begin{align}
I_m(\delta x)&=\frac{\delta_{m,-1}(1-\delta x)+\delta_{m,0}(1+\delta x')+\delta_{m,1}\delta x}{2+\delta x'}\; , \\
I_l(\delta y)&=\frac{\delta_{l,-1}(1-\delta y)+\delta_{l,0}(1+\delta y')+\delta_{l,1}\delta y}{2+\delta y'}\; ,
\end{align}
\end{subequations}
Finally, we sum the image intensities corresponding to all the particles to generate a snapshot (see Fig.~\ref{fig:validation}(a) for an example). Similarly to the experimental setup described in Ref.~\cite{Kurzthaler:2022}, we generate snapshots from data measured at two different magnifications. In particular, we sample $1\times$~magnification data with pixel size $\delta \ell = 6.5\upmu$m, time step $\Delta t=0.02$s, and depth of field $h=400\upmu$m using a simulation box of size $4000\upmu$m$\times4000\upmu$m$\times500\upmu$m.  The small-length-scale data are measured at a higher ($4\times$) magnification. The corresponding parameters are $\delta \ell=1.4\upmu$m, $\Delta t = 0.01$s, and $h=80\upmu$m using a simulation box of size $1000\upmu$m$\times1000\upmu$m$\times160\upmu$m. In both cases the number of pixels is $N_p\times N_p=512\times512$. To mimic the experimental procedure, we then apply the DDM analysis described in Section~\ref{sec:ddm} to extract the ISF from the generated simulation snapshots.

\bibliography{swimmer3d}
\end{document}